# Enhanced Ferroelectric Functionality in Flexible Lead Zirconate Titanate Films with In-Situ Substrate-Clamping Compensation

*Rachel Onn Winestook,[1,2,†] Cecile Saguy,[2,†] Chun-Hao Ma,[3,4,5]*

*Ying-Hao Chu[3,4,5] and Yachin Ivry[1,2]\**

[1]Department of Materials Science and Engineering, Technion – Israel Institute of Technology, Haifa 3200003, Israel.

[2]Solid State Institute, Technion – Israel Institute of Technology, Haifa 3200003, Israel.

[3]Department of Materials Science and Engineering, National Chiao Tung University, Hsinchu 30010, Taiwan.

[4]Institute of Physics, Academia Sinica, Taipei 11529, Taiwan.

[5]Material and Chemical Research Laboratories, Industrial Technology Research Institute, Hsinchu, 31040, Taiwan.

[†]These authors contributed equally to the current work.

[*]Correspondence to: ivry@technion.ac.il.

**Abstract**

Much attention has been given recently to flexible and wearable integrated-electronic devices, with a strong emphasis on real-time sensing, computing and communication technologies. Thin ferroelectric films exhibit switchable polarization and strong electro-mechanical coupling, and hence are in widespread use in such technologies, albeit not when flexed. Effects of extrinsic strain on thin ferroelectric films are still unclear, mainly due to the lack of suitable experimental systems that allow cross structural-functional characterization with in-situ straining. Moreover, although the effects of intrinsic strain on ferroelectric films, e.g. due to film-substrate lattice mismatch, have been investigated extensively, it is unclear how these effects are influenced by external strain. Here, we developed a method to strain thin films homogenously in-situ, allowing functional and structural characterization while retaining the sample under constant straining conditions in AFM and XRD. Using this method, we strained the seminal ferroelectric, $PbZr_{0.2}Ti_{0.8}O_3$ that was grown on a flexible mica substrate, to reduce substrate clamping effects and increase the tetragonality. Consequently, we increased the domain stability, decreased the coercive field value and reduced imprint effects. This method allows also direct characterization of the relationship between the lattice parameters and nanoscale properties of other flexible materials.

**Introduction**

The growing interest in flexible and foldable electronics raises the need for functional materials with suitable characteristics, mainly in the form of thin films. Ferroelectrics are functional materials that exhibit reversible spontaneous polarization, while they are a sub-group of piezoelectrics and hence demonstrate strong electro-mechanical coupling. Ferroelectrics are thus used in a broad range of applications that may benefit from mechanical flexibility, including sensors and medical monitoring devices as well as mobile-phone antennae and non-volatile memory devices. However, characterizing functional properties of thin ferroelectric films at the device-relevant length scale while flexing the material is a great challenge.

Effects of mechanical strain on the piezoresponse and polarization hysteresis loops of thin ferroelectric films were observed first macroscopically (e.g. with optical interference methods), demonstrating clockwise or counterclockwise rotation of the hysteresis loop under tensile and compressive strain respectively.[1–9] Later, local measurements took place by means of piezoresponse force microscopy (PFM).[3] These measurements that were done at the sub-micrometer scale, i.e. relevant for miniaturized devices, revealed an imprint effect or longitudinal shift of the hysteresis loop rather than rotation.[10] For instance, individual PZT capacitors were examined e.g. for 1.5 x 1.5 µm devices of 200-nm thick PZT films grown on Si.[11] The domain distribution and hysteretic behavior of the ferroelectric film were tested with a local probe, by means of piezoresponse force microscopy (PFM) before and after bending the Si substrate by using a holder with a fixed-curvature radius of 30 cm. It was shown that flexing the material results in domains with a single-polarization and a heavily imprinted state, compressive strain inducing positive imprint (hysteresis loop shifts to lower voltages), while tensile strain induces negative imprint (hysteresis loop shifts to the higher voltage values). This observation was then confirmed[12] when a three-point bending stage was used to strain a similar system of PZT on a thinned substrate as well as with dog-bone straining stage.[13] Despite the clear significance of these experiments, rigorous testing of the effects of strain on functionality requires a careful choice of both the flexing method as well as the material examined, mainly because thinned substrate are typically not stable mechanically, while not all substrates can be thinned reproducibly. Moreover, because structural and functional properties are coupled in ferroelectrics, characterizing the hysteresis loop is not sufficient. Rather, the crystallographic structure of the material also needs to be examined under the same strain conditions. Moreover, hysteresis loop is not the only important functional property and also domain stability of the flexed films also has to be characterized.

Among the flexing methods, four-point stressing produces the most homogeneous strain distribution over a long range, enabling characterization of the functional properties independent of the position of the PFM probe as well as under the larger-scale XRD beam. The applied stress ($\sigma$) and hence the induced strain ($\varepsilon$) along a beam according to the torque behavior is described as follows[14,15]:

$$\sigma = \frac{M\delta}{2I} = \frac{3FL_1}{b\delta} \quad (1),$$

where $M = \frac{FL_1}{4}$ is the torque that is applied to the load span $L_1$ (i.e. the distance between the two inner supports of the four stressing points), $F$ is the force, $\delta$ and $b$ are respectively the thickness and width of the bent sample and $I = \frac{b\delta^3}{12}$ is the moment of inertia. The strain is then extracted from Hook's law: $\varepsilon = \frac{\sigma}{E}$ where E is Young's modulus. In four-point bending, the induced torque is constant and the corresponding strain is thus homogeneous along the load span (see Figure 1a). However, applying four-point bending in-situ while performing PFM characterization is a complex task and thus far this method has been used only to characterize ferroelectrics in the bulk form.[16]

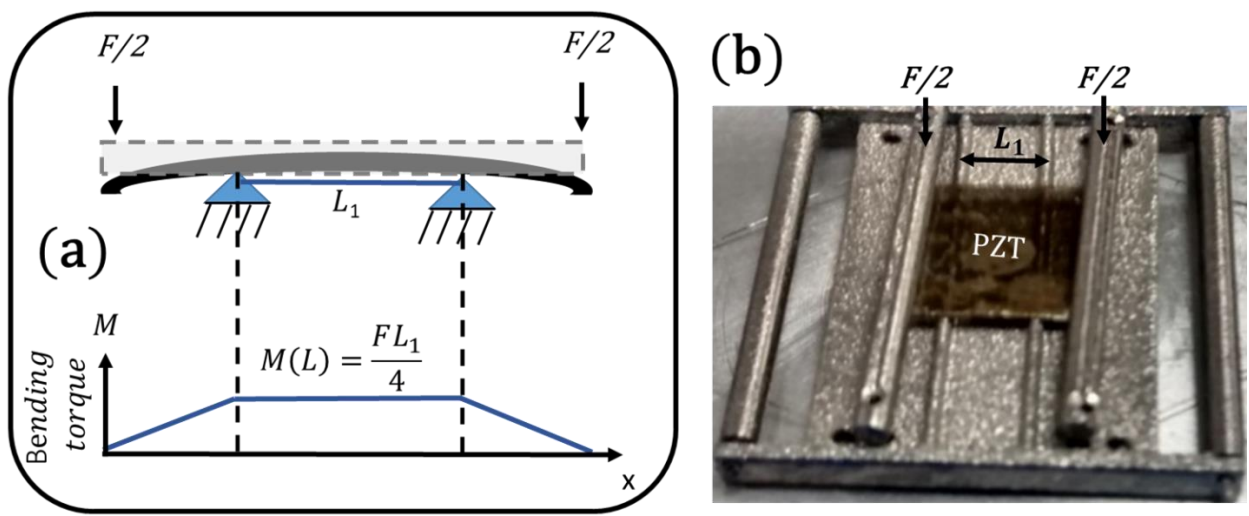

**Figure 1| Tunable homogeneous strain with in-situ four-point bending**. (**a**) Schematic illustration of a four-point bending setup (top) and the resultant homogeneous torque and strain distribution at the load span, i.e. between the inner supports (bottom). (**b**) Optical photo of 1×1 cm$^2$ PZT-SRO-mica in the four-point bending stage that was fabricated by a 3D metallic printer to allow in-situ tunable straining during structural (XRD) and functional (PFM) characterizations (contrast difference in the PZT layer is due to inhomogeneous bottom silver paste spread that is observable through the transparent sample).

In addition to extrinsic strain, thin ferroelectric films experience intrinsic strain that must be considered as well. Thin films of perovskite ferroelectrics exhibit typically inferior functional characteristics with respect to their bulk form because they are inherently strained by the substrate. Significant effort has been put in understanding and controlling the effect of substrate-film lattice mismatch during the film growth on key properties, such as piezoresponse, polarization retention, polarization imprint and coercivity, with an emphasize on the commonly used perovskite ferroelectric, PZT.[13,17–19] It is possible to reduce the effects of substrate clamping by growing thicker films or by lithographically patterning isolated island structures.[5,7,20–24] However, these methods require special treatments and processing steps, which complicates the integration of ferroelectric-based devices in existing technologies.[23] Another strategy that has proved useful for reducing substrate clamping and maintaining the high functional performance of ferroelectrics even as thin films is choosing a substrate with minimal lattice mismatch for the ferroelectric film.[1,25] Nevertheless, because both the film and the substrate are brittle the technological applicability of this strategy is not ideal for flexible devices. Moreover, although the substrate lattice constant can be engineered rather freely to produce a broad range of substrate-film lattice mismatch, this method does not allow strain tunability in a given film, suggesting that direct examination of the effect is limited. Moreover, most commercially relevant ferroelectric films are deposited at high temperature as paraelectrics and they undergo a structural phase transition when being cooled down to the ferroelectric state, so that the lattice matching can be engineered either to the paraelectric or to the ferroelectric state, but not to both. Hence, having a suitable material (films and substrates) and a stressing method that allow in-situ tunability of homogeneous strain is attractive both for understanding the effects of substrate clamping on ferroelectricity as well as for integrating ferroelectrics in flexible and foldable electronic devices. Nevertheless, to-date, experimental realization of the simultaneous extrinsic (flexing) and intrinsic (substrate clamping) effects of strain for the sake of understanding the effects of strain and structure on ferroelectric functionality in a given material is still lacking.

Here, we developed a method for straining a flexible ferroelectric film homogeneously with a four-point bending stage and characterizing both its structural and functional properties in-situ. We used this method to tune the substrate-film lattice mismatch while straining the material in the XRD. We then transferred the sample while maintaining the same strain conditions to perform PFM imaging and spectroscopic characterization. We demonstrated that by compensating the substrate clamping with external strain the ferroelectric functional properties are enhanced.

**Experimental**

The samples investigated in this work are 30-nm thick $PbZr_{0.2}Ti_{0.8}O_3$ (PZT) films grown on a flexible freshly cleaved mica muscovite substrate (30-μm thick) by pulsed laser deposition (PLD).[26] A $CoFe_2O_4$ seeding layer has been used to form a Van de Waals quasi-epitaxy structure (atomic layer of interdiffusion was found, please see details elsewhere[26,27]), while an intermediate layer of 100-nm thick $SrRuO_3$ (SRO) bottom electrode was deposited below the PZT film. The tetragonal PZT composition was chosen to be far away from the morphotropic transition for identifying confidently the effects of structural changes and substrate-film lattice matching on the ferroelectric properties. To apply homogeneous tensile strain on the film, we developed and fabricated a four-point-bending stage that allows in-situ straining while performing nanoscale PFM imaging and spectroscopy and XRD structural characterization as well as transferring the sample between the two, while retaining the same strain conditions (see Figure 1b). This method allowed us to strain the sample during the XRD characterization, and once the obtained strain was as required, the sample was transferred to the PFM while remaining in the holder and subject to the same strain. Hence, in-situ strain tunability was obtained as well as reliable cross-characterization of the structural-functional properties. The bending stage was manufactured using a novel 3D metal-printer (Arcam A2Xm) using $Ti_6Al_4V$ titanium alloy, which has high mechanical strength, even for patterns as small as 0.6 mm as well as high melting and creep temperatures and high electrical conductivity. The method is based on melting of pre-alloyed metal powder particles by an electron beam.

A comparison between the functional properties of the unstrained and strained PZT thin films was done by means of PFM imaging and PFM spectroscopy (Asylum Research, MFP Infinity).[28] The measurements were performed with Si cantilevers coated with titanium silicide, with nominal force constant of 2.4 N·m$^{-1}$ and 70 kHz resonance frequency. The local amplitude and phase hysteresis loops were performed by positioning the PFM tip at various points of the surface and cycling a DC voltage between -10 V and +10 V with 0.1 V increments, superimposed with the AC PFM imaging.[29] Each loop was averaged over at least 85 consecutive cycles. The crystal structure properties were determined with a Rigaku SmartLab 9 kW high-resolution diffractometer. A Cu k$_\alpha$ rotating-anode source at 45 kV tube voltage was used, with a 200-mA tube current as well as a 0-dimension silicon drift detector. Reciprocal space mappings (RSM) were done with a Ge(220)X2 monochromator in parallel-beam geometry. The peaks' positions were found using Rigaku 3D Explore software, where the peaks fit to a 2D Gaussian curve.

**Results and discussion**

The procedure we used to corroborate the structural and functional properties of the same individual samples in unstrained and strained states while applying in-situ external stress was as follows. First, we determined domain stability, coercive field and remanent piezoresponse by PFM imaging and local hysteresis measurements of unstrained samples. Secondly, we used the RSM to measure the lattice parameters of both the PZT and SRO films, allowing us to extract the PZT-SRO lattice mismatch $\left(\frac{a_l - a_s}{a_s}\right)$ as well as the PZT tetragonality $\left(\frac{c_l}{a_l}\right)$, where $a_s$ is the SRO lattice parameter and $a_l$, and $c_l$ are the short- and long- axis PZT lattice parameters. Thirdly, using the bending stage, we varied the PZT-SRO lattice mismatch by tuning the strain in the sample while performing XRD profiling. Finally, once we minimized the lattice mismatch, we transferred the sample to the PFM and characterized the strained samples by both imaging and hysteresis measurements. This iterative process was performed when tuning a given sample controllably in-situ.

**Structure (x-ray and RSM)**

The out-of-plane $\theta - 2\theta$ scans were measured (see Figure SI1), indicating that the major planes parallel to the sample surface are PZT(111) and SRO(111) on mica(001). To determine the crystallographic orientation and lattice parameters we performed an RSM scan. Figure 2a shows asymmetric reciprocal space mapping of PZT(310), PZT(103) and SRO(301) of the unstrained sample. No mica peaks were observed around these reciprocal-space coordinates. The peaks of SRO and PZT were both aligned to $Q_x$, which is the direction parallel to the film in the reciprocal space. That is, the RSM data confirm that the PZT layer was laid in epitaxial registry with the SRO layer. The lattice parameters of both the PZT and the SRO were extracted from the relationship between the inter-planar spacing $d_{hkl}$ and the Miller indices ($h$, $k$ and $l$):

$$\frac{1}{d_{(hkl)}^2} = \frac{h^2+k^2}{a_l^2} + \frac{l^2}{c_l^2} \quad (2),$$

allowing us to extract both the film-substrate lattice mismatch and the PZT tetragonality (for extracting the SRO lattice parameters, we replaced in this expression both $a_l$ and $c_l$ with $a_s$). We then strained the sample in-situ with the four-point bending stage and extracted the corresponding change in lattice parameters by mapping the RSM signal around two different orientations. Figure 2b shows the RSM signal of the strained sample around the PZT(111) orientation (i.e. a symmetric RSM, which allowed us to confirm the hetero-epitaxial mosaic structure of the PZT-SRO layers that are Van der Waals-bonded attached to the mica flakes substrate.[26] Likewise, Figure 2c shows the RSM signal of the same sample, under the same tensile strain, but around the PZT(310) orientation (i.e. asymmetric RSM).

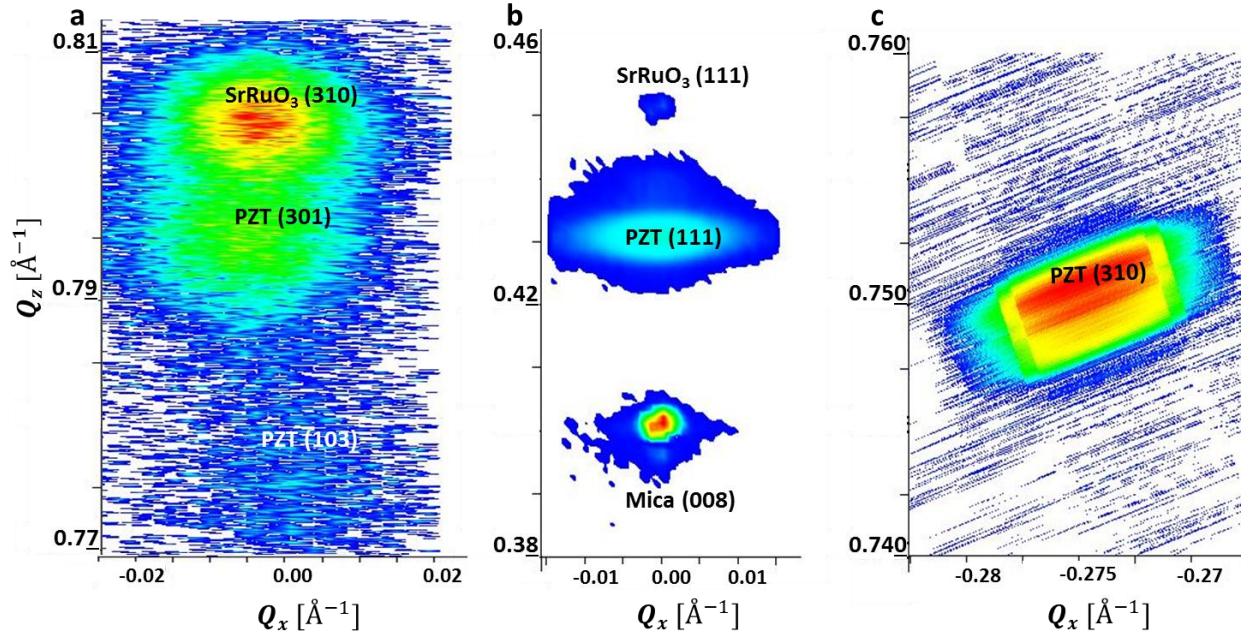

**Figure 2| Reciprocal space mapping of a flexible PZT film without and with strain**. (a) Asymmetric RSM of an unstrained PZT-SRO film on mica. (b) Symmetric and (c) asymmetric RSM of the same material that was now strained in-situ. The position of the {310} and (111) orientations within the reciprocal space axes parallel ($Q_x$) and perpendicular ($Q_z$) to the film allows quantitative calculation of the PZT and SRO lattice parameters at both the unstrained and strained states (see Table 1).

**Table 1| Effects of strain on lattice parameters, film-substrate lattice mismatch and tetragonality**. Lattice parameters of the PZT layer and SRO substrate with and without in-situ strain that were extracted from the corresponding RSM measurements (Figure 2). The reduction of PZT-SRO lattice mismatch and increase in PZT tetragonality in the strained state are also presented.

| PZT film state | $a_l(Å)$ | $c_l(Å)$ | $a_s(Å)$ | PZT-SRO lattice mismatch $\frac{a_l - a_s}{a_s}$ [%] | Tetragonality $\frac{c_l}{a_l}$ |
|---|---|---|---|---|---|
| Unstrained | 3.957 | 4.106 | 3.93 | 0.67 | 1.038 |
| Strained | 3.954 | 4.170 | 3.95 | 0.15 | 1.054 |

The lattice parameters extracted for the PZT and SRO layers of the same sample both unstrained and under strain are given Table 1. Using these parameters, we calculated both the PZT-SRO lattice mismatch and the PZT tetragonality (note that here, the tetragonality is used as a figure of merit to describe the strain in the film). Table 1 also shows that straining the samples reduced the lattice mismatch to less than a quarter of its original values with respect to the unstrained film. Likewise, straining the sample increased the PZT tetragonality by more than 1.5 at %.

**Functionality (PFM imaging and spectroscopy)**

To determine the effects of strain on ferroelectricity we tested both the domain stability as well as the spectroscopic switching parameters. First, we imaged engineered domains (areas scanned with positive and negative voltages that exceed the coercive value) of the unstrained and of the strained sample. The imaging was done by means of amplitude and phase PFM signals of the vertical (out-of-plane) and lateral (in-plane) piezoresponse (typically, proprtional to the polarization) mapping as seen in Figure 3.

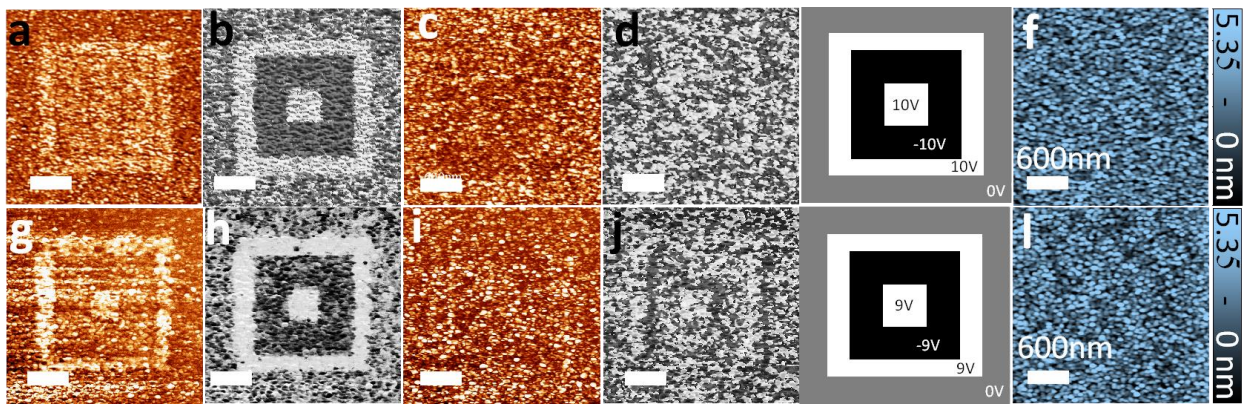

**Figure 3| Polarization distribution of domains in flexible PZT film without and with extrinsic strain immediately after patterning**. (**a**) Out-of-plane PFM amplitude and (**b**) phase signals and the simultaneously imaged (**c**) in-plane PFM amplitude and (**d**) signals showing the polarization distribution in an area that was pre-patterned by applying voltage between the tip and SRO layer as illustrated in the map in (**e**). (**f**) The simultaneously imaged topography of the same area is given as a reference. (**g**) The corresponding PFM out-of-plane amplitude and (**h**) phase as well as (**i**) in plane amplitude and (**j**) phase signals demonstrating the polarization distribution in a pre-

patterned area, while the film experienced in-situ strain. (**k**) The domain patterning map and (**l**) topography of the area of the strained film experiment.

To determine the domain stability, we imaged the relaxation dynamics of the engineered domains (Figures 3b and 3g) over time. The vertical PFM phase image of the relaxing domains in both the strained (Figure 4a-d) and unstrained (Figure 4f-i) show a significant change in the domain relaxation behavior. The unstrained domains relaxed already after about one and a half hours, while the strained sample, with the engineered domains remained stable even after twenty one hours in the strained sample with the reduced PZT-STO lattice mismatch. The native domain distribution of the unstrained and strained samples are given as a reference (Figures 4e and 4j respectively), showing no significant difference between them. To illustrate the clear increase in domain stability for the strained samples, we plotted in Figure 4k the evolution of the percentage of down-polarization domains (dark areas in Figure 4) over time. The native domain distribution of the unstrained and strain films are also given in Figure 4a and 4f, respectively. Figure 4k shows that the native domain were randomly distributed for the unstrained sample and have a preferable 'down' orientation for the strained samples (corresponding to the data points that are marked with 'x').

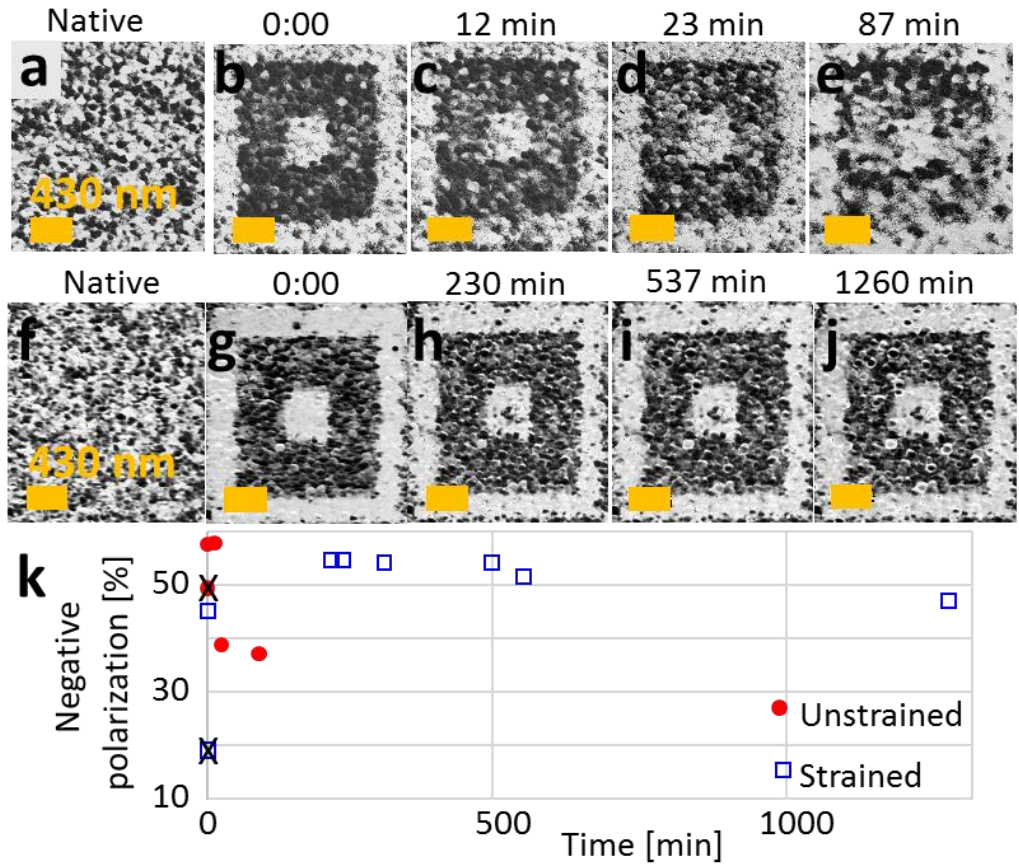

**Figure 4| Domain relaxation in a PZT film with and without external strain**. (a)-(j): Out-of-plane PFM phase signal of the (**a**) native domain distribution in an unstrained film. (**b**) The same area immediately after domain patterning, following the scan map in Figure 3e. (**c**) The domain evolution in this area 12 min, (**d**) 23 min and (**e**) 87 min after the completion of the imaging in (b) showing significant relaxation of the patterned domains. After straining the film (see Table 1), the out-of-plane PFM signal was recorded in a different area to show the (**f**) native domain distribution as well as (**g**) the polarization distribution immediately after domain patterning, following the scan map in Figure 3k. The areas were then scanned repeatedly, demonstrating that the domain distribution was almost unchanged after (**h**) 230 min, (**i**) 537 min and (**j**) 1260 min. The percentage of area with down polarization (dark areas in a-g) a function of time is plotted (**k**) to compare the domain stability in the unstrained and strained scenarios. Filled red circle and the blue square correspond to relaxation in the unstrained and flexed sample, respectively. Empty data points at time '0' designated with 'x' represent the native domain distribution.

Next, to complete the examination of the effects of strain on the functional properties of the ferroelectric material, we measured the switching parameters of the hysteresis loop in the PZT film. Local hysteresis loop measurements were done in a set of 8×7 points spread in a 25×25 $\mu m^2$ area. At each point, four consecutive switching cycles were performed. For examining the effect of strain on the hysteresis loops, the entire experiment was done both for the unstrained and strained sample with the same experimental conditions (same cantilever, voltage-sweeping conditions etc.) with only a few hours gap between the measurements of the unstrained and strained sample.

Figure 5a shows the average butterfly (PFM amplitude) hysteresis loops of the unstrained (red circles) and strained (blue squares) states of the same film. The negative and positive coercive voltages measured from the amplitude butterfly hysteresis-loops are -5.75 V and 6.70 V for the unstrained state, and -3.60 V and 5.4 V for the strained state. That is, straining the ferroelectric films gives rise to reduction in the coercive voltage value with respect to the native unstrained state with the larger film-substrate lattice mismatch. We also compared the piezoresponse hysteresis loops (amplitude times cosine the phase of the PFM signal) of the unstrained and strain states of the ferroelectric film as seen in Figure 5b. Here, we clearly see that the straining the sample not only has increased the saturation values of the piezoresponse signal (which is proportional to the polarization) and increased the area in the hysteresis loop, but also made the hysteresis loop more symmetric than the unstrained state with respect both to the remanent piezoresponse and coercive voltage. The parameters extracted from the switching spectroscopy are given in Table 2. We should note that Figures 5a and 5b are the average of the last three (out of four) consecutive measurements that were done in each point. For the sake of completeness, the average of the first switching cycle for all the 8×7 points of the unstrained and strained cases are given for both the butterfly (PFM amplitude signal, Figure 5c) and piezoresponse (PFM amplitude times the cosine of the PFM phase signal, Figure 5d) are given as a reference.

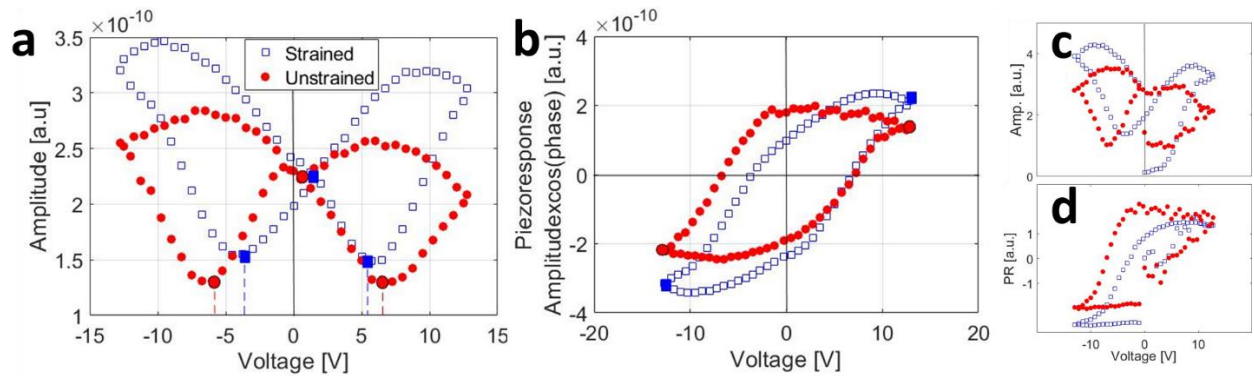

**Figure 5| Hysteresis loop measurements of the PZT film both unstrained and under in-situ strain**. (**a**) Butterfly (PFM amplitude signal) and (**b**) piezoresponse (amplitude times cosine the phase signal) hysteresis loops measured in the unstrained (full red circles) PZT and under in-situ strain (empty blue squares). Each hysteresis curve is an average of the last three (out of four) switching cycles from 56 different locations. The averaged (**c**) butterfly and (**d**) piezoresponse hysteresis loops of the first switching measurements in these 56 points is given as a reference.

**Table 2| Functional properties of unstrained and strained PZT**. The switching parameters that are extracted from the averaged hysteresis loops (see Figure 5) of the PZT film both without and with external strain. The positive and negative coercive values, positive and negative saturation piezoresponse (which is proportional to the saturation polarization) and remanent voltage were extracted (see highlighted data points in Figure 5).

|  | Unstrained | Strained |
|---|---|---|
| **Positive coercive voltage [V]** | 6.70 | 5.4 |
| **Negative coercive voltage [V]** | -5.75 | -3.60 |
| **Positive piezoresponse saturation [a.u.]** | 1.68 | 2.120 |
| **Negative piezoresponse saturation [a.u]** | -2.21 | -3.320 |
| **Remanent voltage [V]** | 0.77 | 1.00 |

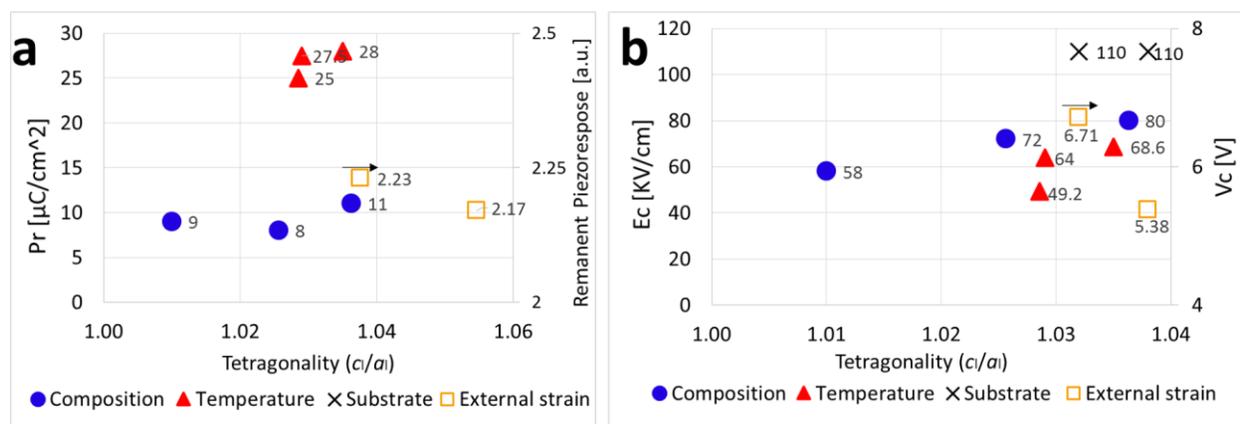

**Figure 6|** Functional properties of PZT films as a functional of their tetragonality: comparison with literature.[30–33] (a) Remanent polarization and remanent piezorespose vs. tetragonality. (b) Coercive field and coercive voltage vs. tetragonality. (See Table SI1 for details).

Our results show a clear relationship between the structure and functionality of ferroelectrics by means of direct observations. Reducing the film-substrate lattice mismatch between the PZT film and the SRO sample is accompanied by significant increase in stabilization of switched ferroelectric domains (retention). Reduction of this lattice matching is also accompanied by reduction of the asymmetry of the ferroelectric hysteresis loop (imprint and surface charging). Straining the sample led also to another change in the structural properties – increase in the PZT tetragonality, which in turn was accompanied by an increase of the remanent piezoresponse as well as the area integrated within the hysteresis loop. We suggest that these observations are in agreement with the existing literature related to the effects of either lattice matching engineering (by means of varying the substrates)[1,33–35] or in-situ bending[11,12,36] on the hysteresis loop and stability of switched domains. However, the in-situ strain tunability during structural and functional characterization allows bridging between these two effects of substrate clamping and flexing in ferroelectrics, i.e. between the effects of intrinsic and extrinsic strain. The observed change in the ferroelectric properties under flexing conditions of the PZT films may open some technological applications for ferroelectrics as flexible electronics. We should note though that the effects of flexing on the ferroelectric properties may vary for different PZT compositions. For instance, existing literature suggests that PZT near the morphotropic transition may exhibit a larger degree of stability in the ferroelectric properties when flexed.[26] To assess the effects of extrinsic strain reported here, we compared it to the effect of intrinsic strain for the switching properties. Figure 6 shows literature data of the dependence of remanent polarization as well as coercive field on tetragonlity for strain obtained by varying the substrate, the PZT composition or the PZT temperature (raw data are given in Table SI1). These data is compared to the dependence of remanent piezoresponse and coercive voltage as a function of tunable tetragonality, which is reported here. This comparison shows that trend of the dependence of these values on the tetragonality varies between the extrinsic strain and the intrinsic strain. The exact origin of this opposite trend is still not completely known to us, mainly with respect to the coercive value. A possible explanation is that as mentioned above, in addition to change in tetragonlity, the strain changed also the lattice-substrate lattice mismatch, which may play a role in these two values. However, a comparison that accounts for both changes in tetragonality and substrate clamping is not straightforward, while not all data do not yet exist in the literature. We therefore encourage further experimental and theoretical investigation of different ferroelectric materials and compositions for flexible-electronic purposes. Finally, we hope our work will encourage further examinations of the relationship between structure and nanoscale functionality of materials that are subject to in-situ tunable strain.

**Acknowledgements**

This research was supported by a grant from the Ministry of Science & Technology, Israel & the Ministry of Science and Technology of Taiwan. The Technion group acknowledges also financial support from the Zuckerman STEM Leadership Program. The authors would like to thank Mr. Garry Muller from the Israeli Institute of Metals for assistance with the 3D printing of the 4-point bending stage as well as Mr. Shimon Cohen and Mr. Alon Hendler Avidor for earlier discussion regarding the design of this stage. Moreover, we thank Dr. Maria Koifman and Prof. Semën Gorfman for their assistance with the XRD and RSM characterizations and data interpretation.

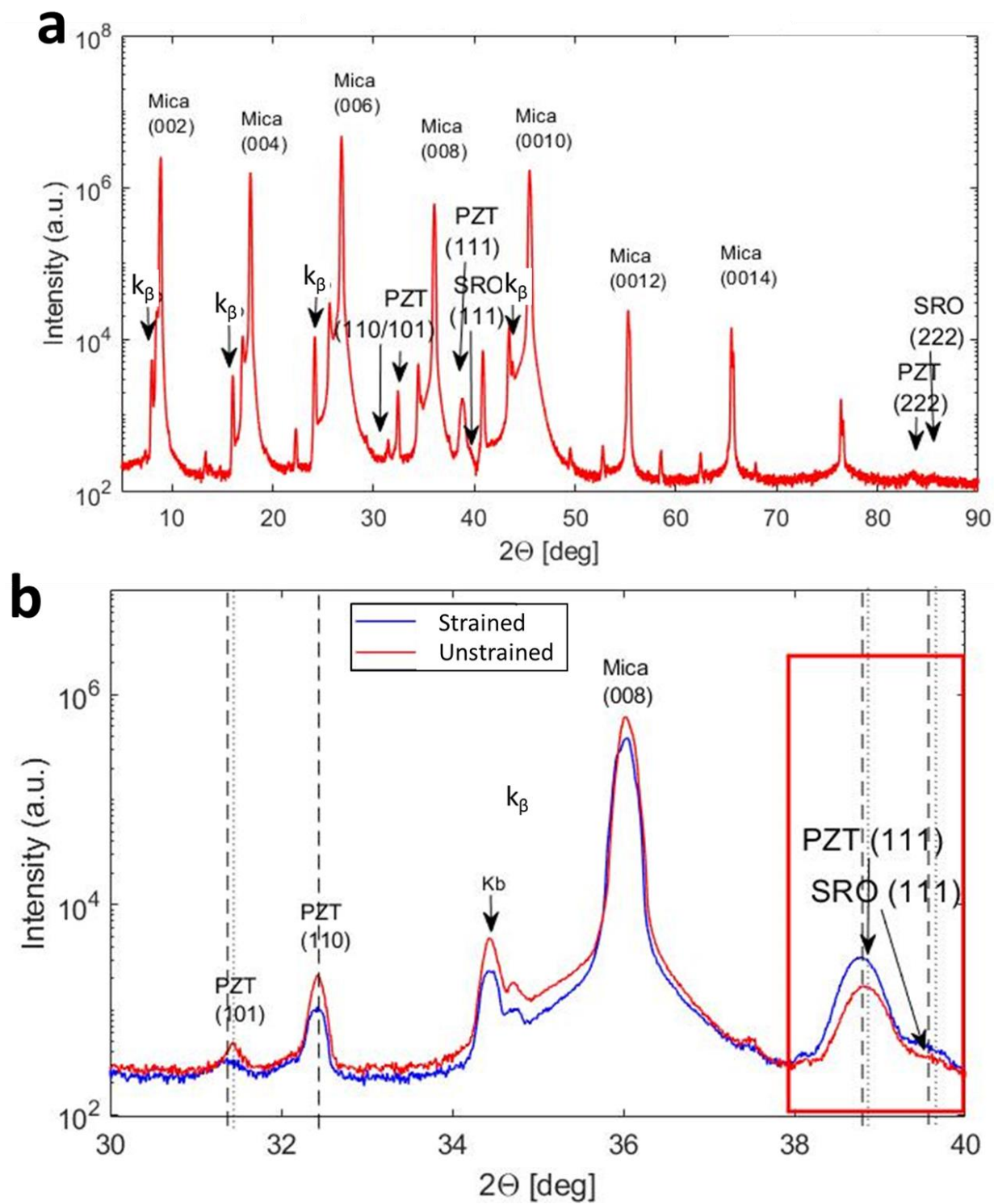

**Figure SI1|** (**a**) Broad and (**b**) zoom-in range $\theta - 2\theta$ x-ray scans of the unstrained (red) and strained (blue) PZT-SRO-mica heterostructure.

Table SI1| Structure-functionality relationship as extracted from this work, in comparison to the existing literature, following Figure 6.

|  |  | $a_l$ [Å] | $c_l$ [Å] | Tetragonality | Ec [kV/cm] | Vc [V] | Pr [uC/cm^2] |
|---|---|---|---|---|---|---|---|
| **Composition [Zr %wt]**[30] | 0.1 | 3.965 | 4.109 | 1.036 | 80 |  | 11 |
|  | 0.2 | 3.98 | 4.082 | 1.025 | 72 |  | 8 |
|  | 0.3 | 4.005 | 4.045 | 1.010 | 58 |  | 9 |
| **Temperature [°C]** [31,32] | 25 | 4 | 4.14 | 1.035 | 68.55 |  | 28 |
|  | 100 | 4.01339 | 4.13 | 1.029 | 64 |  | 27.5 |
|  | 140 | 4.0161 | 4.1308 | 1.028 | 49.21 |  | 25 |
| **Substrate**[33] | STO (100) | _ | _ | 1.038 | 110 |  | 22 |
|  | MGO(100) | _ | _ | 1.032 | 110 |  | 17 |
| **External strain** | Unstrained | 3.957 | 4.106 | 1.037 |  | 5.38 | 2.23 * |
|  | Strained | 3.954 | 4.17 | 1.054 |  | 6.71 | 2.17* |